\documentclass[longauth,letter]{aa}  
\usepackage{graphicx}
\usepackage{natbib}

\newcommand{\HII}{H\,{\sc ii}}

\newcommand{\NeII}{[Ne\,{\sc ii}]}

\usepackage{txfonts}
\begin{document}
\title{{\it Herschel} observations in the ultracompact HII region Mon~R2\thanks{Herschel is an ESA space observatory with 
science instruments 
provided by European-led Principal Investigator consortia and with 
important participation from NASA.} }
\subtitle{Water in dense Photon-dominated regions (PDRs)}
\author{A.~Fuente\inst{1}, O.~Bern\'e\inst{2}, J.~Cernicharo\inst{3},
J.R.~Rizzo\inst{3}, M.~Gonz\'alez-Garc\'{\i}a\inst{4}, J.R.~Goicoechea\inst{3}
P.~Pilleri\inst{5,6}, V.~Ossenkopf\inst{7,8}, M. Gerin\inst{9}, R.~G\"usten\inst{10},
M.~Akyilmaz\inst{7}, A.O.~Benz\inst{11}, F.~Boulanger\inst{12}, S. Bruderer\inst{11},
C.~Dedes\inst{11}, K.~France\inst{13}, S. Garc\'{\i}a-Burillo\inst{14}, A.~Harris\inst{15},
C.~Joblin\inst{5,6}, T.~Klein\inst{10}, C.~Kramer\inst{4}, F.~Le Petit\inst{16}, 
S.D.~Lord\inst{17}, P.G.~Martin\inst{13}, J.~Mart\'{\i}n-Pintado\inst{3},B.~Mookerjea\inst{18},
D.A.~Neufeld\inst{19}, Y.~Okada\inst{7}, J. Pety\inst{20}, T.G.~Phillips\inst{21}, M.~R\"ollig\inst{7},
R.~Simon\inst{7}, J.~Stutzki\inst{7}, F.~van der Tak\inst{8,22},D.~Teyssier\inst{23}, A. Usero\inst{14},
H.~Yorke\inst{24}, K.~Schuster\inst{20}, M. Melchior\inst{25}, A.~Lorenzani\inst{26}, R. Szczerba\inst{27},
M.~Fich\inst{28}, C.~McCoey\inst{28,29}, J. Pearson\inst{24}, P.~Dieleman\inst{30}}

\institute{
Observatorio Astron\'omico Nacional (OAN), Apdo. 112, 28803 Alcal\'a de Henares (Madrid), Spain
\and
Leiden Observatory, Universiteit Leiden, P.O. Box 9513, NL-2300 RA Leiden, The Netherlands 
\and
Centro de Astrobiolog\'ia, CSIC-INTA, 28850, Madrid, Spain
\and
Instituto de Radio Astronom\'ia Milim\'etrica (IRAM), Avenida Divina Pastora 7, Local 20, 18012 Granada, Spain
\and
Universit\'e de Toulouse, UPS, CESR, 9 avenue du colonel Roche, 31062 Toulouse cedex 4, France
\and
CNRS, UMR 5187, 31028 Toulouse, France
\and
I. Physikalisches Institut der Universit\"at 
zu K\"oln, Z\"ulpicher Stra\ss{}e 77, 50937 K\"oln, Germany
\and
SRON Netherlands Institute for Space Research, P.O. Box 800, 9700 AV 
Groningen, Netherlands
\and
LERMA, Observatoire de Paris, 61 Av. de l'Observatoire, 75014 Paris, France 
\and
Max-Planck-Institut f\"ur Radioastronomie, Auf dem H\"ugel 69, 53121, Bonn, Germany
\and
Institute for Astronomy, ETH Z\"urich, 8093 Z\"urich, Switzerland
\and
Institut d'Astrophysique Spatiale, Universit\'e Paris-Sud, B\^at. 121, 91405 Orsay Cedex, France
\and
Department of Astronomy and Astrophysics, University of Toronto, 60 St. George Street, Toronto, ON M5S 3H8, Canada
\and
Observatorio Astron\'omico Nacional (OAN), Alfonso XII, 3, 28014 Madrid, Spain
\and
Astronomy Department, University of Maryland, College Park, MD 20742, USA
\and
Observatoire de Paris, LUTH and Universit\'e Denis Diderot, Place J. Janssen, 92190 Meudon, France
\and
IPAC/Caltech, MS 100-22, Pasadena, CA 91125, USA
\and 
Tata Institute of Fundamental Research (TIFR), Homi Bhabha Road, Mumbai 400005, India
\and 
Department of Physics and Astronomy, Johns Hopkins University, 3400 North Charles Street, Baltimore, MD 21218, USA
\and
Institut de Radioastronomie Millim\'etrique, 300 Rue de la Piscine, 38406 Saint Martin d'H\'eres, France
\and
California Institute of Technology, 320-47, Pasadena, CA  91125-4700, USA
\and
Kapteyn Astronomical Institute, University of Groningen, PO box 800, 9700 AV Groningen, Netherlands
\and
European Space Astronomy Centre, Urb. Villafranca del Castillo, P.O. Box 50727, Madrid 28080, Spain
\and
Jet Propulsion Laboratory, 4800 Oak Grove Drive, Pasadena, CA 91109  U.S.A.
\and
Institut für 4D-Technologien, FHNW, 5210 Windisch, Switzerland 
\and
Osservatorio Astrofisico di Arcetri-INAF- Largo E. Fermi 5 I-50100 Florence, Italy 
\and
N. Copernicus Astronomical Center, Rabianska 8, 87-100, Torun, Poland 
\and
Department of Physics and Astronomy, University of Waterloo, Waterloo, ON Canada N2L 3G1 
\and
University of Western Ontario, Dept. of Physics \& Astronomy, London,
Ontario, Canada N6A 3K7
\and
SRON Netherlands Institute for Space Research, Landleven 12, 9747 AD Groningen
}
\authorrunning {A.~Fuente, et al.} 
\titlerunning{HIFI observations of the UCHII region Mon~R2}

\abstract
{
Mon R2, at a distance of 830 pc, is the only ultracompact \HII\ region (UC \HII)  where the photon-dominated region (PDR) between 
the ionized gas and the molecular cloud can be resolved with {\it Herschel}. Therefore, it is an excellent laboratory to study the  
chemistry in extreme PDRs (G$_0$$>$10$^5$ in units of Habing field, n$>$10$^6$~cm$^{-3}$).}
{Our ultimate goal is to probe the physical and chemical conditions in the PDR
around the UC \HII\ Mon~R2.}
{
HIFI observations of the abundant compounds $^{13}$CO, C$^{18}$O, o-H$_2$$^{18}$O, HCO$^+$, CS,
CH, and NH have been used to derive the physical and chemical conditions in the PDR, in particular
the water abundance. The modeling of the lines has been done with the Meudon PDR code and the 
non-local radiative transfer model described by Cernicharo et al. (2006).
}
{The $^{13}$CO, C$^{18}$O, o-H$_2$$^{18}$O, HCO$^+$ and CS observations are well described assuming that the emission
is coming from a dense ($n$=5$\times$10$^6$~cm$^{-3}$, N(H$_2$)$>$10$^{22}$~cm$^{-2}$) layer of molecular gas
around the \HII\ region. Based on our o-H$_2$$^{18}$O observations, we estimate an o-H$_2$O abundance 
of $\approx$2$\times$10$^{-8}$. This is the
average ortho-water abundance in the PDR. Additional H$_2$$^{18}$O and/or water lines are required to derive the
water abundance profile. A lower density envelope ($n$$\sim$10$^5$~cm$^{-3}$, N(H$_2$)=2-5$\times$10$^{22}$~cm$^{-2}$) is responsible for the
absorption in the NH 1$_1$$\rightarrow$0$_2$ line. The emission of the CH ground state triplet is coming from both regions 
with a complex and self-absorbed profile in the main component.
The radiative transfer modeling shows that the $^{13}$CO and HCO$^+$ line profiles are consistent with an expansion of the molecular gas
with a velocity law, $v_e$ =0.5$\times$(r/R$_{out}$)$^{-1}$~km~s$^{-1}$, although the expansion velocity
is poorly constrained by the observations presented here.
}
{We determine an ortho-water abundance of $\approx$2$\times$10$^{-8}$ in Mon~R2. 
Because shocks are unimportant in this region and our estimate is based on H$_2$$^{18}$O observations 
that avoids opacity problems, this is probably the most accurate estimate of the water abundance in PDRs thus far.}
 

\maketitle

\section{Introduction}

Ultracompact (UC) \HII\ regions constitute one of the earliest phases in the formation of a massive
star and are characterized by extreme physical and chemical conditions (G$_0$$>$~10$^5$ in units of Habing field
and n$>$~10$^6$~cm$^{-3}$). Their understanding is important for distinguishing the different processes in the massive
star formation process and because they can be used as a template for other extreme photon-dominated regions (PDRs) such us
the surface layers of circumstellar disks and/or the nuclei of starburst galaxies.
The UC \HII\  Mon~R2 is the only one that can be resolved with {\it Herschel}.

Mon R2 is a nearby (d=830~pc; Herbst \& Racine 1976) complex star forming region. It
hosts a UC \HII\ region near its center, powered by the infrared source 
Mon~R2~IRS1 (Wood \& Churchwell 1989). The molecular content of this region has been the 
subject of several observational studies. The huge CO bipolar outflow (Meyers-Rice \&
Lada 1991), $\sim$15$'$ long (=3.6 pc) is a relic of the formation of the B0V star
associated to IRS1 (Massi, Felli, \& Simon 1985; Henning, Chini, \& Pfau 1992) and strong shocks
are currently not at work in this region (Bern\'e et al. 2009). 
Previous molecular observations (Giannakopoulou et al. 1997;
Tafalla et al. 1997; Choi et al. 2000; Rizzo et al. 2003,2005)
showed that the UC \HII\ region is located inside a cavity and bound by a dense molecular ridge. 
The peak of this molecular ridge (hereafter, MP) is located at an offset (+10$"$,$-$10$"$) relative to the peak of
the ionized gas (hereafter, IF). The molecular hydrogen column density
toward the MP is 2$-$6$\times$10$^{22}$~cm$^{-2}$. The detection of the reactive ions CO$^+$ and
HOC$^+$ showed a dense photon-dominated region (PDR) surrounding
the UC \HII\ region (Rizzo et al. 2003, 2005).
Recent Spitzer observations probed the thin molecular gas layer 
(n=4$\times$10$^5$ cm$^{-3}$, N(H$_2$)=1$\times$10$^{21}$~cm$^{-2}$) with T$_k$=574($\pm$20)~K in between 
the ionized gas and the dense molecular gas traced by previous millimeter observations (Bern\'e et al. 2009). 
All these components are schematically shown in Fig. 1. 

\onlfig{1}{
\begin{figure}
\includegraphics{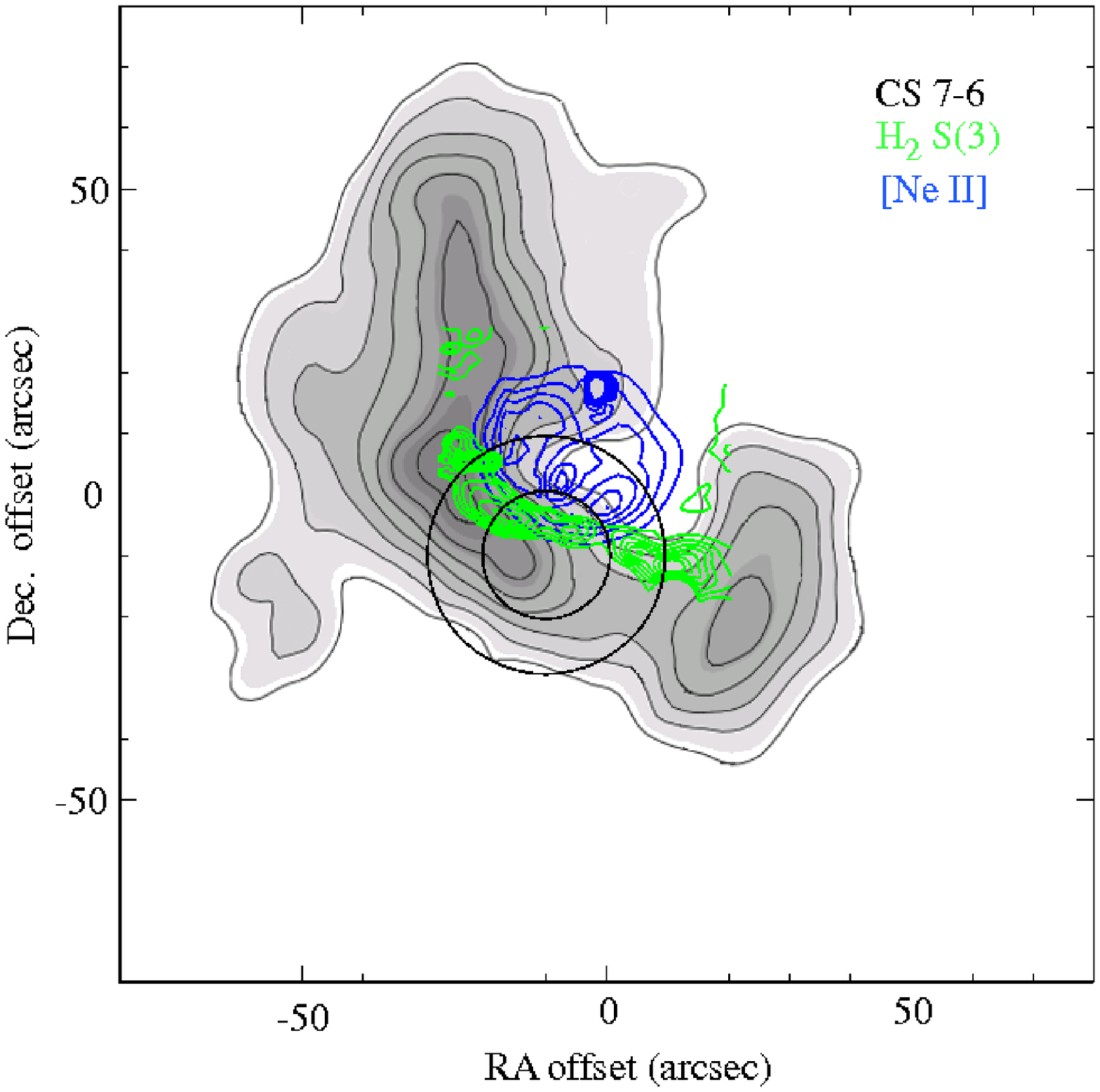}
\vspace{9.5cm}
      \caption{Overview of the PDR associated with Mon~R2. Gray scales represent the CS 7$\rightarrow$6 line emission map from Choi et al. (2000).
The blue contours represent the intensity of the \NeII\ line (0.02--0.1 erg s$^{-1}$ cm$^{-2}$ sr$^{-1}$ in linear steps),
and the green contours the intensity of the H$_2$ S(3) rotational line at
9.7 $\mu$m (1.5--4.5 $\times$10$^{−4}$ erg s$^{-1}$ cm$^{-2}$ sr$^{-1}$ in linear steps). 
The circles indicate the {\it Herschel} beam in band 1a (40$"$) and band 4a (21$"$) centered on the molecular peak (MP).
}
         \label{Fig 1}
 \end{figure}
}

\setcounter{figure}{1}
\begin{figure}
\includegraphics{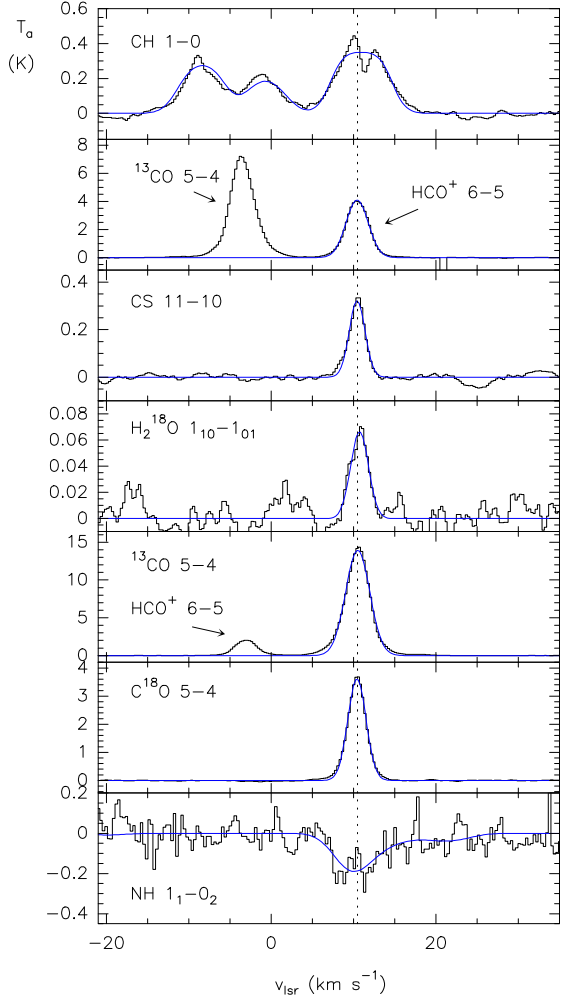}
\vspace{13.5cm}
      \caption{HIFI spectra toward the MP in Mon~R2. Note that the $^{13}$CO 5$\rightarrow$4 and HCO$^+$ 6$\rightarrow$5 lines appear close in velocity. 
This is an artifact of the DSB observations. The HCO$^+$ 6-5 line is observed in the LSB and the
$^{13}$CO 5$\rightarrow$4 line, in the USB. The dashed line indicates v$_{lsr}$=10.5~km~s$^{-1}$.
}
         \label{Fig 1}
 \end{figure}

\section{Observations}

The observations were made with the HIFI instrument onboard
{\it Herschel} (Pilbratt et al. 2010, de Graauw et al. 2010) during the Priority Science 
Phase~2 in the frequency switch (FSW) observing mode
with a reference position at the offset (+10$'$,0). Two receiver settings 
were observed, one in Band 1a with the WBS centered at 536.066GHz (LSB) 
and the other in Band 4a with the WBS centered at 971.800~GHz (LSB). These
settings were observed toward both positions, IF [RA=06h07m46.2s,
DEC=-06$^\circ$23$'$08.3$''$ (J2000)] and MP [RA=06h07m46.87s,DEC=-06$^\circ$23$'$18.3$''$ (J2000)]. In this letter 
we present the observations toward the MP because the study of the spatial distribution 
of the molecular tracers is postponed for a forthcoming paper. The data were reduced using 
HIPE 3.0 pipeline. The adopted intensity scale was antenna temperature. In addition to {\it Herschel}
data, we used the HCO$^+$ (1$\rightarrow$0), HCO$^+$ (3$\rightarrow$2), CS (2$\rightarrow$1),
$^{13}$CO (1$\rightarrow$0), $^{13}$CO (2$\rightarrow$1), C$^{18}$O (1$\rightarrow$0) and C$^{18}$O 
(2$\rightarrow$1) observed with the IRAM 30m telescope.

\section{Results}
We detected the CH 1$\rightarrow$0, HCO$^+$ 6$\rightarrow$5,
CS 11$\rightarrow$10, o$-$H$_2^{18}$O 1$_{10}$$\rightarrow$1$_{01}$, $^{13}$CO 5$\rightarrow$4, C$^{18}$O 5$\rightarrow$4, and NH 1$_1$$\rightarrow$0$_2$
lines in the MP (see Fig. 2). We stress that this is the first detection of the rarer water isotopologue o$-$H$_2^{18}$O toward a
spatially resolved PDR. All the lines except NH 1$_1$$\rightarrow$0$_2$ were detected in emission. Gaussian fits are shown in Table 1 and Fig. 2. 
We fitted the three components of the CH 1$\rightarrow$0 line assuming the same excitation
temperature and estimated that the opacity of the main component is $>$3. In this case, the central velocity is not well 
determined because of the self-absorption and the flattened profiles produced by the large opacities. 
The NH 1$_1$$\rightarrow$0$_2$ line was tentatively detected ($\sim$3$\sigma$) in absorption. This line is composed of 10 components
and we fitted all of them assuming the same excitation temperature. Our fit shows that 
the NH line is optically thin. Because the individual components are not resolved with a
linewidth of $\sim$4~km/s, the individual line parameters are uncertain. This detection needs to be confirmed.

\begin{table}
\caption{Summary of HIFI observations}
\label{tab_observation}
{\scriptsize
\begin{center}
\begin{tabular}{llcccc} \hline 
\multicolumn{1}{l}{Line} & \multicolumn{1}{c}{Freq.} & 
\multicolumn{1}{c}{T$_a$$\times$$\tau$} & \multicolumn{1}{c}{v$_{lsr}$} &
\multicolumn{1}{c}{$\Delta v$} & \multicolumn{1}{c}{$\tau$} \\ 
\multicolumn{1}{l}{} & \multicolumn{1}{c}{(GHz)} & 
\multicolumn{1}{c}{(K)} & \multicolumn{1}{c}{(km s$^{-1}$)} &
\multicolumn{1}{c}{(km s$^{-1}$)} & \multicolumn{1}{c}{} \\ 
\hline
 CH  &   536.761 & 1.81(0.21)  &  10.9(0.1)  &  4.9(0.3) &  3.6(0.6) \\
 NH  &   974.478 & -0.24(0.07) &   9.9(0.4)  &  4.9(0.8) &  0.1(0.7) \\ \hline 
\multicolumn{1}{l}{Line} & \multicolumn{1}{c}{Freq} &
\multicolumn{1}{c}{Area} & \multicolumn{1}{c}{v$_{lsr}$} &
\multicolumn{1}{c}{$\Delta v_{lsr}$} & \multicolumn{1}{c}{T$_{MB}$} \\ 
\multicolumn{1}{l}{} & \multicolumn{1}{c}{(GHz)} &
\multicolumn{1}{c}{(K$\times$km~s$^{-1}$)} & \multicolumn{1}{c}{(km s$^{-1}$)} &
\multicolumn{1}{c}{(km s$^{-1}$)} & \multicolumn{1}{c}{(K)} \\ 
\hline
HCO$^+$     & 535.061  &  19.4(4.0)  & 10.4(0.3)   &  3.2(0.8)  &  5.7 \\
CS          & 538.688  &  1.09(0.04) & 10.5(0.1) &  2.2(0.1)  & 0.4 \\
o$-$H$_2^{18}$O & 547.676  &  0.22(0.03) & 10.7(0.1)   &  2.4(0.3)  & 0.08 \\
$^{13}$CO   & 550.926  &  51.1(2.0)  & 10.6(0.1) &  3.4(0.2)  &  14.0 \\
C$^{18}$O   & 548.830. &  14.0(0.4)  & 10.5(0.1) &  2.6(0.1)  &  5.0 \\
\hline
\end{tabular}
\end{center}
}
\end{table}

\subsection{Line profiles}
\label{sect_profiles}
Different velocity components can be distinguished in this region. The ambient cloud is centered at 
v$_{lsr}$=10.5$\pm$1~km~s$^{-1}$. A large scale molecular outflow is associated with IRS~1 (Meyers-Rice \& Lada, C.J. 1991;
Tafalla et al. 1997).
The wings observed in the emission profiles of the HCO$^+$ 1$\rightarrow$0 and 3$\rightarrow$2 lines and in the $^{13}$CO 2$\rightarrow$1 line 
(velocity ranges [0,6]~km~s$^{-1}$ and [4,14]~km~s$^{-1}$) are associated with the molecular outflow.
The HCO$^+$ 6$\rightarrow$5, $^{13}$CO 5$\rightarrow$4 and C$^{18}$O 5$\rightarrow$4 lines are centered at a velocity of $\sim$10.4~km~s$^{-1}$.
In Fig. 3 we compare the HIFI lines with the low rotational lines of the same species observed with the IRAM 30m telescope. The angular resolution
of the IRAM data was degraded to match those of {\it Herschel}. The line profiles of the low rotational lines of HCO$^+$ and $^{13}$CO are self-absorbed at 
red-shifted velocities. For C$^{18}$O, there is a perfect match between the profiles of the the J=5$\rightarrow$4 and J=2$\rightarrow$1 lines. 
The profiles of the CS J=2$\rightarrow$1 and J=11$\rightarrow$10 lines also match perfectly.

The HCO$^+$ 6$\rightarrow$5, CS 11$\rightarrow$10, $^{13}$CO 5$\rightarrow$4 and C$^{18}$O 5$\rightarrow$4 lines
have characteristic linewidths of 2--3~km~s$^{-1}$. The largest linewidths observed in the NH 1$_1$$\rightarrow$0$_2$ 
and CH lines ($\sim$ 5~km~s$^{-1}$) indicate that these 
lines are tracing a different, probably more diffuse component. As argued in Sect 4.1,
the NH absorption is very likely caused by the cold and lower density envelope surrounding the UC \HII\ region. 
The CH 1$\rightarrow$0 line is seen in absorption and emission suggesting that CH is present in the low density envelope 
and the dense PDR. 

\setcounter{figure}{2}
\begin{figure}
\includegraphics{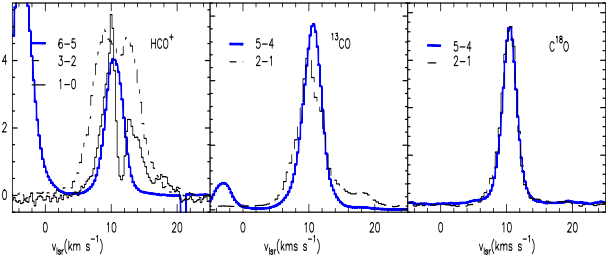}
\vspace{3.7cm}
      \caption{Comparison of the profiles of the HIFI and IRAM 30m spectra toward the MP in Mon R2. The high angular resolution 30m spectra were
degraded to match the angular resolution of {\it Herschel} data (40$"$). The $^{13}$CO 1$\rightarrow$0 and  C$^{18}$O 1$\rightarrow$0 lines are not included because we observed them with a poor spectral resolution (1.5~MHz$\sim$7~km~s$^{-1}$). 
}
         \label{Fig 1}
 \end{figure}


\subsection{Molecular column densities}
In Table 2 we present the estimated molecular column densities. We used the rotational diagram technique to estimate the rotation temperature
and the column density of the CO isotopologues and CS.
This technique gives an average rotation temperature and total column density providing that the emission is optically thin. For optically thick
lines, we obtained a lower limit to the true column density and rotation temperature (this is very likely the case for $^{13}$CO). For CH and NH
we assumed a reasonable value of the rotation temperature.
Previous molecular observations at millimeter wavelengths showed
the existence of gas with densities up to n=4$\times$10$^6$~cm$^{-3}$ and T$_k$$\approx$50~K in the molecular ridge (Choi et al. 2000; Rizzo et al. 2003, 2005). Bern\'e et al. (2009) derived a density, n=4$\times$10$^5$~cm$^{-3}$ and a gas kinetic temperature of $\sim$570~K
for a thin gas layer around the \HII\ region on basis of the H$_2$ rotational lines.
These two n-T pairs of values can produce the rotation temperatures derived from our observations. However, the gas layer at 570~K cannot account for the observed line intensities. For instance, assuming a standard C$^{18}$O abundance of 2$\times$10$^{-7}$, this hot layer will produce a C$^{18}$O 5$\rightarrow$4 line intensity of T$_b$$\sim$0.09~K, i.e. 2\% of the observed value. Therefore, the emission of the observed lines is mainly coming from the dense (n=4$\times$10$^6$~cm$^{-3}$) molecular ridge.

We detected the o$-$H$_2^{18}$O 1$_{10}$$\rightarrow$1$_{01}$ line in the MP. Using the non-local radiative transfer
code of Cernicharo et al. (2006) and assuming the densities and temperatures prevailing in the dense molecular ridge
(T$_k$=50~K, n(H$_2$)=4$\times$10$^6$~cm$^{-3}$), we obtained an excitation temperature for the ground state transition of 
o$-$H$_2^{18}$O of $\sim$8.5~K. With this low excitation temperature,
we need N(o$-$H$_2^{18}$O)= 2.7$\times$10$^{12}$~cm$^{-2}$ to fit our observations. Assuming a $^{18}$O/ $^{16}$O ratio 
of $\sim$500, this implies an ortho-water abundance of $\sim$2$\times$10$^{-8}$. The excitation temperature is not
very sensitive to the gas kinetic temperature. To assume a gas kinetic temperature as high as 500~K would increase the 
excitation temperature by a factor of 2, and decrease the estimated o$-$H$_2^{18}$O column density by a factor of 10. 
To assume a lower hydrogen density would be more critical, because the excitation temperature would drop
to very low values ($<$6~K), and the line would become very weak and optically thick which would prevent
any good estimate of the o$-$H$_2^{18}$O column density.
The water abundance estimated in Mon~R2 is similar to that obtained toward the Orion Bar by Olofsson et al. (2003) using ODIN observations.
In that case, the spatial resolution of the ODIN observations did not allow the authors to resolve the dense PDR. 
Moreover, their estimate was based on observations of the ground state transition of the main water isotopologue, which is optically thick. 
The agreement between the two measurements could therefore be fortuitous. Recent SPIRE observations of the Orion Bar have provided 
an upper limit to the water abundance of a few 10$^{-7}$ (Habart et al. 2010).

To interpret the observed NH 1$_1$-0$_2$ line absorption, a knowledge
of the continuum level is needed. Unfortunately, the line was observed in FSW mode,
which removes the continuum. Hence, the continuum level had to be estimated from previous measurements.
In particular, Dotson et al. (2010) measured an intensity of
280~Jy/beam using the CSO telescope (beam=20$"$) at 350~$\mu$m
(=857~GHz). Assuming a spectral index $\beta$ of $\sim$3, we estimate a
continuum flux of 411 Jy at 974 GHz. This value corresponds to
a brightness temperature of 0.8~K in the HIFI beam. Taking into
account the uncertainty in $\beta$ and the different beams of
{\it Herschel} and CSO, we estimate that the accuracy of the continuum intensity at 974~GHz
is about a factor of 2.
Assuming T$_{ex}$=10\,K, and an intrinsic linewidth of 4\,km~s$^{-1}$, the
line opacity inferred from the observed
absorption feature implies a NH column density of
(1--5)$\times$10$^{13}$ cm$^{−2}$
(a factor $\sim$1.5 higher if T$_{ex}$=20\,K).
Taking $\sim$5$\times$10$^{22}$~cm$^{-2}$ as an upper limit for N(H$_2$)
(the NH absorption likely arises in a external layer of more diffuse gas),
the NH abundance would be greater than (1.0-0.2)$\times$10$^{-9}$. These abundances are comparable
to the NH abundance first inferred by ISO toward Sgr B2 (Cernicharo et al. 2000, Goicoechea et al. 2004).

\begin{table}
\caption{Molecular column densities$^*$}
\label{tab_observation}
\begin{center}
\begin{tabular}{llcl} \hline 
\multicolumn{1}{l}{Mol} & \multicolumn{1}{c}{T$_{rot}$} & 
\multicolumn{1}{l}{Observed} & \multicolumn{1}{c}{PDR Model}  \\ 
\multicolumn{1}{l}{} & \multicolumn{1}{c}{(K)} & 
\multicolumn{1}{c}{(cm$^{-2}$)} & \multicolumn{1}{c}{(cm$^{-2}$)}  \\ 
\hline
$^{13}$CO       & 27   & 4.7 10$^{16}$         &   3.5 10$^{16}$$^b$    \\
C$^{18}$O       & 34   & 6.9 10$^{15}$         &   3.5 10$^{15}$$^b$    \\
H$^{13}$CO$^+$  & 20$^c$   & 1.7 10$^{12}$$^c$     &   9.3 10$^{11}$$^b$    \\
CS              & 25   & 6.3 10$^{13}$         &  6.7 10$^{13}$     \\
o$-$H$_2^{18}$O     & 8.5$^a$   & 2.7 10$^{12}$    &  2.9 10$^{12}$$^{b,d}$      \\
CH              & 19$^a$    & 5.6 10$^{13}$    &  3.3 10$^{14}$    \\
NH              & 10-20$^a$  & 1-5 10$^{13}$   &  5.1 10$^{11}$      \\ \hline
\end{tabular}
\end{center}
\indent
$^*$ The rotational diagrams used the 30m and HIFI lines.

\noindent
$^a$ Assumed rotational temperature.

\noindent
$^b$ Assuming $^{12}$C/$^{13}$C=50 and $^{16}$O/$^{18}$O=500

\noindent
$^c$ From Rizzo et al. (2005) 

\noindent
$^d$ Assuming an ortho-to-para ratio of 3.
\end{table}

\section{Discussion}
\subsection{Chemical model}
We explored the possibility of explaining the molecular abundances observed in Mon~R2 in
terms of PDR chemistry. To do this, we used the updated version of the Meudon PDR code (Le Petit et al. 2006,
Goicoechea \& Le Bourlot 2007). As input parameters we used a plane-parallel slab with a thickness of 10~mag and n=4$\times$10$^6$~cm$^{-3}$, 
which is illuminated from the left side with a field of G$_0$=5$\times$10$^5$ and from the right side with the standard
interstellar UV field (G$_0$=1). Elemental abundances were: He (0.1), O (3.19$\times$10$^{-4}$), N (7.5$\times$10$^{-5}$),C (1.32$\times$10$^{-4}$),
S (1.86$\times$10$^{-5}$), Fe (1.5$\times$10$^{-8}$). The model does not include isotopic fractionation, 
we estimated the abundances of the isotopologues
assuming $^{12}$CO/$^{13}$CO=50 and $^{16}$O/$^{18}$O=500. The obtained column densities are shown in Table 2 and
the abundance profiles are plotted in Fig. 4. We got an excellent agreement between observations and model predictions
for the CO isotopologues, HCO$^+$, H$_2$O, and CS. The model
successfully predicts the observed average water abundance. The water abundance is far from uniform across the PDR,
with a peak of $\sim$10$^{-6}$ at an extinction of $\sim$~1~mag and a minimum value $<$10$^ {-10}$ (see Fig. 4).
The observation of high excitation lines of water will allow us to derive the water abundance profile and further constrain the chemical modeling.
The model falls short by more than one order of magnitude however to predict the NH column density.
The NH abundance is $<$10$^{-9}$ all across the PDR (see Fig. 4).
As argued below, this is very likely due to the main contribution to the observed NH line coming from the low density envelope that is protected from the UV radiation.
We ran the code for a low density (n$\approx$4$\times$10$^4$~cm$^{-3}$) plane-parallel layer of 50~mag illuminated with G$_0$=1 and obtain a NH column density of $\sim$3$\times$10$^{13}$~cm$^{-2}$ in better agreement with the observations. This low density envelope would also contribute to the CH ground state line since the CH column density should be $\sim$ a few 10$^{14}$~cm$^{-2}$.
For this reason, we have a very complex CH profile, with the line optically thick and self-absorption in the main 
component. However, the contribution of the low density envelope to the other observed HIFI lines would be negligible because
of the low density and low gas kinetic temperature. For instance, the intensity of the $^{13}$CO 5$\rightarrow$4 line would be $\sim$0.1~K.

\subsection{Kinematics of the region}
The systematic change in the line profiles of the rotational lines of HCO$^+$ and $^{13}$CO can be understood in terms of the velocity structure of the molecular core. Although the complete modeling of the source is beyond the scope of this paper, (most of the HIFI data are still to come), we made a preliminary model to gain insights into the kinematics of this source.
The UC \HII\ region is expected to expand because of the different pressure of the ionized and molecular gas (Jaffe et al. 2003; Rizzo et al. 2005). We modeled the HCO$^+$ and $^{13}$CO lines using the non-local radiative transfer code of Cernicharo et al. (2006). In the model, we adopted the physical structure derived in Sect. 4.1 (a spherical envelope with inner radius R$_{in}$=0.08~pc, an innermost layer with T$_k$=50~K, n=4$\times$10$^6$~cm$^{-3}$ and a radial thickness of 0.0006~pc, and an external envelope with the temperature decreasing as T$\propto$R$^{-0.5}$, a constant density of n=1$\times$10$^5$~cm$^{-3}$ and a thickness of 0.16~pc). For the dust temperature and opacity, we adopted the values derived by Thronson et al. (1980). As a first approximation we assumed constant $^{13}$CO and HCO$^+$ abundances of 2$\times$10$^{-6}$ and 10$^{-9}$ respectively and only varied the velocity law. We were able to reproduce the line intensities and the trend observed in the line profiles assuming an expansion velocity law v$_e$= 0.5$\times$(R/R$_{out}$)$^{-1}$~km~s$^{-1}$ (R$_{out}$=0.24~pc) and a turbulent velocity of 2~km~s$^{-1}$ (see Fig. 5). Because of the hole in the molecular core, the maximum expansion velocity is 1.5~km~s$^{-1}$, which is similar to the turbulent velocity. For this reason, the back and front parts of the envelope are radiatively coupled producing the self-absorbed profiles at red-shifted velocities. The shapes of the line profiles are dominated by the turbulent velocity with little effect of the small expansion velocity, which is poorly constrained.

\onlfig{4}{
\begin{figure}
\includegraphics{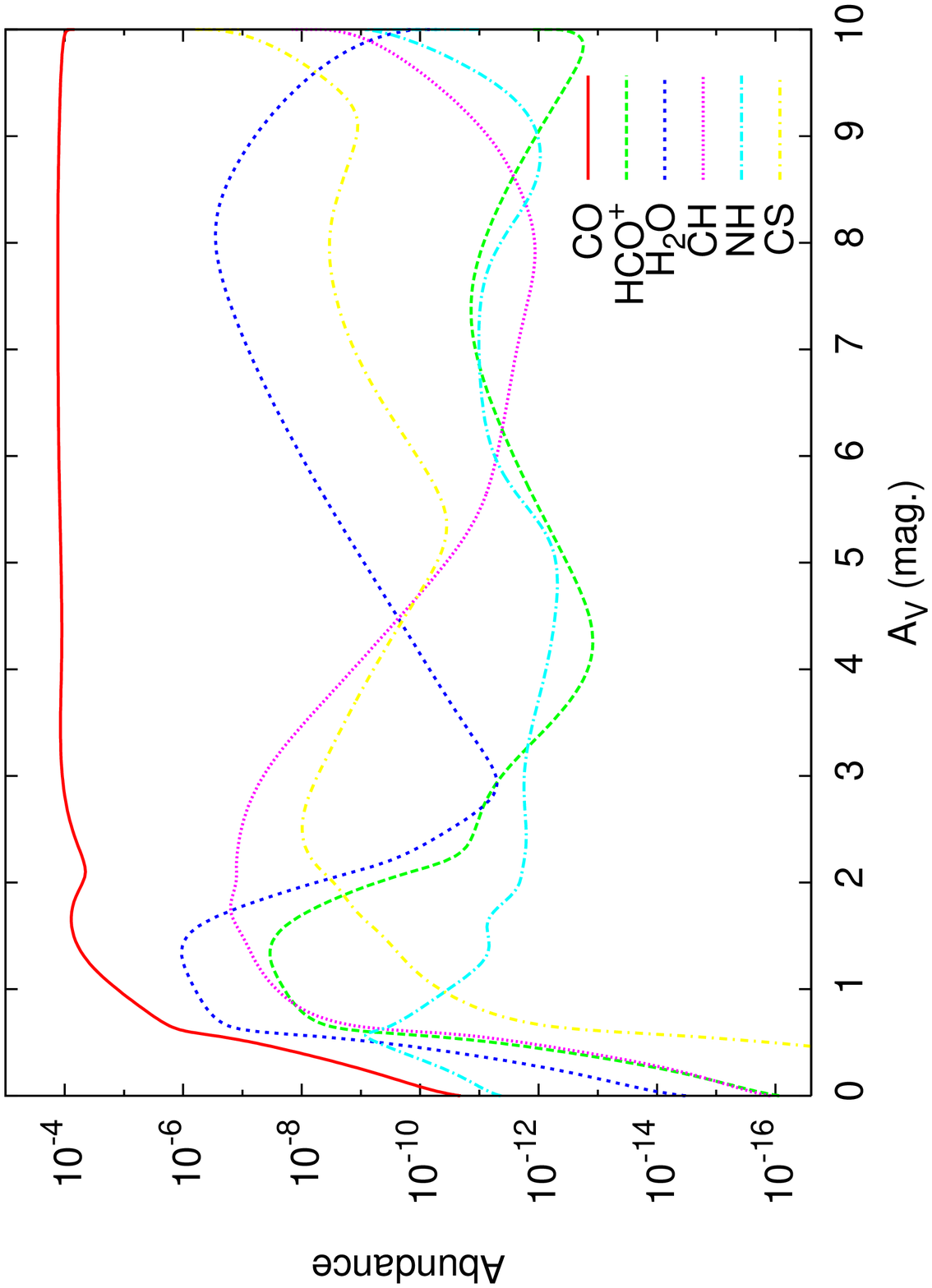}
\vspace{8.0cm}
      \caption{Results of the chemical modeling of the dense PDR surrounding the UCHII region Mon~R2}
         \label{Fig 1}
 \end{figure}
}

\setcounter{figure}{4}
\begin{figure}
\includegraphics{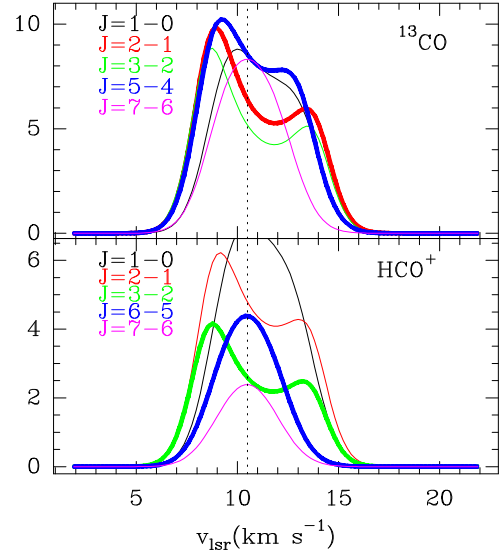}
\vspace{8.0cm}
      \caption{Predicted line profiles for the $^{13}$CO and HCO$^+$ rotational lines using the model described in Sect 4.2. 
}
         \label{Fig 1}
 \end{figure}

\section {Conclusions}
We present the first HIFI observations toward the UC \HII\ region Mon~R2. Detections of the
$^{13}$CO 5$\rightarrow$4, C$^{18}$O~5$\rightarrow$4, o$-$H$_2$$^{18}$O 1$_{10}$$\rightarrow$1$_{01}$, HCO$^+$ 
6$\rightarrow$5, CS 11$\rightarrow$10, NH 1$_1$-0$_2$, and CH 1$\rightarrow$0 lines are reported.
The emission/absorption of all these molecules is well explained assuming the the molecular core is
composed of a dense (n=5$\times$10$^6$~cm$^{-3}$) PDR layer of gas surrounded by a lower
density UV protected envelope. The modeling of the $^{13}$CO and HCO$^+$ line profiles is consistent with the molecular gas being expanding with an expansion velocity law, v$_e$=0.5$\times$(r/R$_{out}$)$^{-1}$~km~s$^{-1}$.
Based on our o$-$H$_2$$^{18}$O 1$_{10}$$\rightarrow$1$_{01}$ observations,
we estimate an ortho-water abundance of $\sim$2$\times$10$^{-8}$. Because shocks are unimportant in this region and our estimate is based on the rarer isotopologue 
observations that avoids opacity problems, this water abundance estimate is probably the most accurate in PDRs thus far.

\begin{acknowledgements}
HIFI has been designed and built by a consortium of institutes and university departments from across
Europe, Canada and the United States under the leadership of SRON Netherlands Institute for Space
Research, Groningen, The Netherlands and with major contributions from Germany, France and the US.
Consortium members are: Canada: CSA, U.Waterloo; France: CESR, LAB, LERMA, IRAM; Germany:
KOSMA, MPIfR, MPS; Ireland, NUI Maynooth; Italy: ASI, IFSI-INAF, Osservatorio Astrofisico di Arcetri-
INAF; Netherlands: SRON, TUD; Poland: CAMK, CBK; Spain: Observatorio Astron\'omico Nacional (IGN),
Centro de Astrobiolog\'{\i}a (CSIC-INTA). Sweden: Chalmers University of Technology - MC2, RSS \& GARD;
Onsala Space Observatory; Swedish National Space Board, Stockholm University - Stockholm Observatory;
Switzerland: ETH Zurich, FHNW; USA: Caltech, JPL, NHSC. This paper was partially supported by Spanish MICINN
under project AYA2009-07304 and within the program CONSOLIDER INGENIO 2010, under grant ”Molecular Astrophysics: 
The Herschel and ALMA Era – ASTROMOL” (ref.: CSD2009-00038).

\end{acknowledgements}

\end{document}